\documentclass[twocolumn,showpacs,amsmath,amssymb]{revtex4}

\usepackage{dcolumn}
\usepackage{bm}
\usepackage{epsfig}
\usepackage{amsfonts}

\begin{document}


\title{Statistics of 3-dimensional Lagrangian turbulence}

\author{Christian Beck}
\affiliation{
School of Mathematical Sciences \\ Queen Mary, University of
London \\ Mile End Road, London E1 4NS, UK}
\email{c.beck@qmul.ac.uk}
\homepage{http://www.maths.qmul.ac.uk/~beck}

\date{\today}

\vspace{2cm}

\begin{abstract}
We consider a superstatistical model for
a Lagrangian tracer particle in a high-Reynolds number
turbulent flow.
The analytical model predictions are in excellent agreement with
recent experimental data for flow between counter-rotating disks.
In particular, the predicted Lagrangian scaling exponents
$\zeta_j$ agree well with the
measured exponents reported in [X. Hu et al., PRL {\bf 96},
114503 (2006)].
The model also correctly predicts the shape of acceleration
probability densities, correlation functions, statistical
dependencies between components and explains
the fact
that enstrophy lags behind dissipation.

\end{abstract}


\date{\today}          
\maketitle

The full understanding of the statistical properties of fluid
turbulence still remains a challenging problem in theoretical
physics. In recent years there has been some experimental
progress in measuring the statistical properties of single tracer
particles advected by turbulent flows
\cite{boden1,boden2,boden3,boden4,boden5,boden6,pinton1,pinton2,pinton3}.
These experiments, as well as direct numerical simulations (DNS)
\cite{yeung1,yeung2,yeung3,bif}, have significantly enhanced our
knowledge of the stochastic properties of Lagrangian turbulence (LT).
A variety of interesting experimental results has recently been
published, in particular for the probability densities of
accelerations \cite{boden1,boden2}, velocity differences
\cite{pinton1}, correlation functions \cite{boden3,pinton2},
conditional expectations \cite{boden3,boden4,pinton3}, the
Lagrangian scaling exponents $\zeta_j$
\cite{pinton2,pinton3,boden5} and the corresponding $f(\alpha)$
spectra obtained by Legendre transformation \cite{boden6}. Many
of these new experimental data confirm early DNS results obtained in
\cite{yeung1}.

The exact theory of LT based on the Navier-Stokes equation still
very much lags behind the experimental progress. Hence it is
important to develop simple theoretical models that provide an
explanation for the most important statistical features of 3-d LT.
Previous models have been succesfull in explaining e.g.\ the 1-d
measured acceleration statistics but fail to explain the recent
experimental data for the 3-d statistics \cite{boden3} or the new
data for the Lagrangian scaling exponents \cite{boden5,boden6}.
It is thus important to develop theoretical models that explain
not just one but {\em all} of the above  measured LT phenomena
with sufficient precision at the same time, using a consistent
set of parameters.

In this letter we will
introduce such a model, which for the first time
simultaneously reproduces the measured 3-d statistics, correlation functions,
statistical dependencies between components and scaling exponents.
We will carefully
compare its theoretical predictions with the available
experimental data, obtaining excellent agreement. Our model is a
natural and physically plausible extension of previous LT
models based on superstatistical stochastic differential
equations (SDEs)
\cite{prl,cohen,reynolds,euro,physica-d,aringazin,lee}. Here the term
`superstatistics' \cite{cohen} means that there is a
superposition of several stochastic processes, a fast one as
given by the original SDE and a slow one for the parameters of
the SDE, which are regarded as slowly varying random variables
describing the changing environment of the Lagrangian tracer
particle. Our model approximates
the high-Reynolds number limit of a
superstatistical extension of the Sawford
model \cite{sawford, pope, euro}, and is more
refined than previous models \cite{prl, euro, reynolds} by taking into account both a fluctuating
energy dissipation and a fluctuating enstrophy surrounding the
test particle.

We obtain predictions for
Lagrangian scaling exponents $\zeta_j$ (and, by Legendre
transform, for multifractal spectra) that are in very good
agreement with the recent measurements of the Bodenschatz group
\cite{boden5,boden6}. The agreement seems to be better than for
other models, e.g.\ the Lagrangian multifractal turbulence models
compared with the data in \cite{boden6}. At the same time our
model correctly reproduces the measured probability densities of
single velocity difference and acceleration components as well as
those of the modulus, it describes correctly the fact that the
three acceleration components are not statistically independent,
it gives the correct conditional acceleration variance, and it
explains the fact that the correlation function for single
acceleration components decays rapidly whereas that of the
modulus decays slowly. Finally, the model also explains why the
fluctuating enstrophy lags behind the fluctuating energy
dissipation, as numerically observed by Yeung and Pope
\cite{yeung1} and experimentally by Zeff et al. \cite{zeff}. To
the best of our knowledge, our model is the first LT
model that simultaneously achieves all this.

To introduce the model, let us denote the velocity of a Lagrangian
tracer particle embedded in the turbulent flow by $\vec{v}(t)$. We
are interested in velocity differences on given time scales
$\tau$, i.e.\ the quantity $\delta
\vec{v}(t)=\vec{v}(t)-\vec{v}(t+\tau)$, which is directly measured
in various experiments. To obtain a compact notation, in the
following we write $\vec{u}(t):=\delta \vec{v}(t)$. Let us
consider a linear superstatistical SDE for $\vec{u}(t)$ of the
following form:
\begin{equation}
\dot{\vec{u}}=-\Gamma  \vec{u} + \Sigma  \vec{L} (t) \label{1}
\end{equation}
Here $\vec{L} (t)$ is a rapidly fluctuating stochastic process
representing force differences in the liquid on a fast time scale,
and $\Gamma$ and $\Sigma$ are 3$\times$3 matrices. We approximate
$\vec{L}(t)$ by Gaussian white noise. $\Gamma (t)$ and $\Sigma
(t)$ are matrix-valued stochastic processes which evolve on a
much larger time scale than $\vec{L} (t)$.

Generally, the class of superstatistical models described by
eq.~(\ref{1}) is quite large. A first example was introduced in
\cite{prl} and further developed in \cite{reynolds,euro}. Let us
here consider a more refined model that takes into account two
very important facts: A fluctuating local energy dissipation rate
of the environment surrounding the test particle and a fluctuating
local enstrophy (rotational energy). We thus consider as a
special case of eq.~(\ref{1}) the local dynamics
\begin{equation}
\dot{\vec{u}}=-\gamma \vec{u}+B \vec{n} \times \vec{u}+\sigma
\vec{L}(t). \label{2}
\end{equation}
We assume that $\gamma$ and $B$ are constants, but the
noise strength $\sigma$ and the unit vector $\vec{n}$ describing
a temporary rotation axis of the particle evolve
stochastically on a large time scale $T_\sigma$ and
$T_{\vec{n}}$, respectively.
$T_\sigma$ is of the same order of magnitude as the integral time
scale $T_L$, whereas $\gamma^{-1}$ is of the same order of
magnitude as the Kolmogorov time scale $\tau_{\eta}$. From the
above one gets $T_\sigma \gamma \sim T_L/\tau_{\eta} \sim
R_\lambda >>1$, i.e.\ the time scale separation between the slow
and the fast processes, which is a necessary condition for the
superstatistics approach to work \cite{BCS}, increases
proportional to the Taylor scale Reynolds number $R_\lambda$.
The time scale $T_{\vec{n}}>>
\tau_\eta$
describes the average life
time of a region of given vorticity surrounding the test particle.

As it is customary in statistical physics, we define
a parameter $\beta:=2\gamma/\sigma^2$, which in equilibrium
statistical mechanics corresponds to the inverse temperature,
whereas in superstatistical turbulence models \cite{prl,euro,physica-d}
it is a formal parameter related to
a suitable power $\epsilon^\kappa$ of the fluctuating energy dissipation rate $\epsilon$.
In the following we adopt the choice $\kappa =-1$, i.e.
$ \beta^{-1} \sim \nu^{1/2} \langle \epsilon \rangle^{-1/2} \epsilon$,
where $\nu$ is the kinematic viscosity and $\langle \epsilon \rangle$ the
average energy dissipation.
To further specify our
superstatistical model we still have to fix the probability
density of the stochastic process $\beta(t)$, which, motivated by
the cascade picture of turbulence and previous successful models
\cite{reynolds,euro,boden4,BCS,K62,castaing}, is assumed to be
close to a lognormal
distribution
\begin{equation}
f(\beta) = \frac{1}{\beta s \sqrt{2\pi}}\exp\left\{ \frac{-(\log
\frac{\beta}{m})^2}{2s^2}\right\}. \label{Logno}
\end{equation}
Here $m$ and $s$ are mean and variance parameters. The average
$\beta_0$ of the above log-normal distribution is given by
$\beta_0=m\sqrt{w}$ and the variance by $\sigma^2=m^2w(w-1)$,
where $w:= e^{s^2}$.


On a time scale $t$ satisfying $\gamma^{-1} <<t <<T_\sigma$ the
probability density of a single component $u_x$ of the tracer
particle described by eq.~(\ref{2}) is given by the Gaussian
\begin{equation}
p(u_x|\beta)=\sqrt{\frac{\beta}{2\pi}}e^{-\frac{1}{2}\beta u_x^2}.
\end{equation}
Note that this result is independent of $B$ and $\vec{n}$ \cite{vKa}.
In the long-term run the variance
of this Gaussian will fluctuate since $\sigma$ fluctuates. Hence we get a superposition of
Gaussians with different variance parameter $\beta^{-1}$, i.e.
the marginal stationary distribution of our superstatistical
system is given by
\begin{equation}
p_{u_x}(u_x)
=\frac{1}{\sqrt{2\pi}}\int_0^\infty \beta^{1/2}f(\beta)
e^{-\frac{1}{2}\beta u_x^2}d\beta, \label{su}
\end{equation}
no matter what the value of $B$ is.
This formula,
with $f(\beta)$ being the lognormal distribution,
is in excellent agreement with experimentally measured histograms
\cite{boden1,boden2,pinton1}.
An
example is shown in Fig.~1.
One obtains good fittings of the data in \cite{pinton1} for all
time scales $\tau$, with
$w\sim (\tau /\tau_\eta)^{-0.4}$.
\begin{figure}[ht]
\includegraphics[scale=0.8]{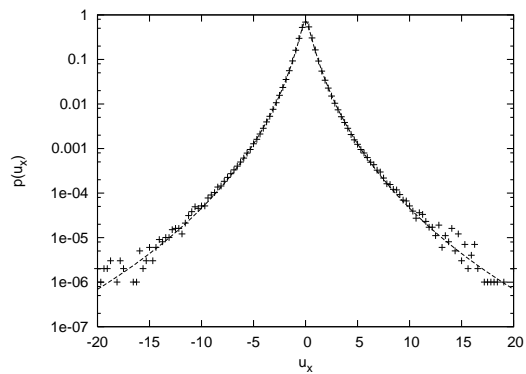}
\caption{Experimentally measured probability distribution
$p(u_x)$ at smallest time scales (data from \cite{pinton1} for
$\tau \approx 0.15$ ms) and a fit by eq.~(\ref{su}) and
(\ref{Logno}) with $s^2=1.8$.}
\end{figure}
For very
small $\tau$ the acceleration of the particle is given by $a_x=u_x/\tau$ and hence by transformation
of random variables $p_{a_x}(a_x)=\tau p_{u_x}(u_x)$, thus
\begin{equation}
p_{a_x}(a_x) = \frac{\tau}{2\pi s }\int_0^\infty d\beta \;
\beta^{-1/2} \exp\left\{ \frac{-(\log
\frac{\beta}{m})^2}{2s^2}\right\} e^{-\frac{1}{2}\beta \tau^2 a_x^2} .
\label{10}
\end{equation}
This formula is in good agreement with the
results presented in \cite{boden1,boden2}
($s^2=3.0$, see \cite{euro} for details).

Let us now check what type of Lagrangian scaling exponents
$\zeta_j$ for velocity increments $u$ are predicted by our
superstatistical model.
From eq.~(\ref{su}) and eq.~(\ref{Logno}) one obtains the moments
\begin{equation}
\langle u^j \rangle =
(j-1)!! m^{-\frac{j}{2}} w^{\frac{1}{8}j^2}. \label{super}
\end{equation}
The notation $(j-1)!!$ stands for a product of all
odd positive integers up to $j-1$.
Assuming simple scaling laws of the form
$m\sim \tau^a$, $w\sim \tau^b$,
where $a$ and $b$ are so far arbitrary real numbers,
eq.~(\ref{super}) implies $\langle u^j\rangle \sim
\tau^{-a\frac{j}{2}+b\frac{1}{8}j^2}\sim \tau^{\zeta_j}$. Hence
\begin{equation}
\zeta_j=-\frac{a}{2}j+\frac{b}{8}j^2.
\end{equation}
It is often assumed that the Lagrangian exponent $\zeta_2$ is
equal to 1.
From $\zeta_2=1$ we get $a=\frac{1}{2}b-1$ thus
\begin{equation}
\zeta_j=(\frac{1}{2}+\lambda^2)j-\frac{1}{2}\lambda^2 j^2, \label{scalex}
\end{equation}
where, following the notation of Mordant et al. \cite{pinton2} we
defined a positive parameter $\lambda^2$ by
$\lambda^2=-\frac{1}{4}b$. Note that we get a formula analogue to
Kolmogorov's K62 theory \cite{K62}, however the difference is that
this formula is directly applicable to the Lagrangian dynamics, it
needs not to be transformed from an Eulerian to a Lagrangian
frame (as it was done in \cite{chevillard,boden6} for multifractal models).
Our formula (\ref{scalex}) is in very  good
agreement with the experimental data presented in \cite{boden6},
see Fig.~2.
\begin{figure}[ht]
\includegraphics[scale=0.8]{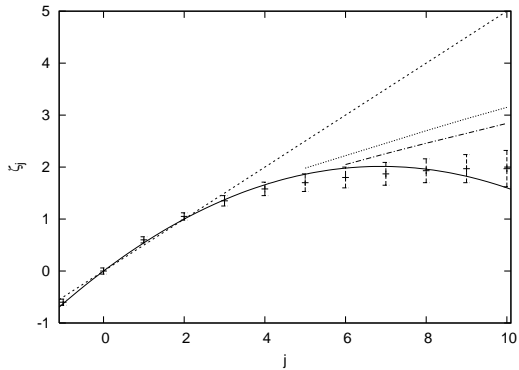}
\caption{Lagrangian scaling exponents $\zeta_j$ as measured by Xu
et al. \cite{boden6} and as predicted by the superstatistical
model for $\lambda^2=0.085$ (solid line). Other model predictions
are also shown (from top to bottom): No intermittency at all (dashed line),
Lagrangian version
of She-L\'{e}v\^{e}que model \cite{she, boden6} (dotted line) and
multifractal model of Chevillard et al. \cite{chevillard, boden6}
(dashed dotted line).}
\end{figure}
The agreement is better than for the other theoretical models
compared with the data in \cite{boden6}. Whereas all our
predicted exponents are within the error bars of the experimental data,
the predictions of the other models are
outside the experimental error bars for $j\geq 5$ (dotted and dashed-dotted lines in Fig.~2).

We may also proceed to
the multifractal turbulence spectra $D(h)$ defined by
$D(h)=\inf_j (hj+1-\zeta_j)$
by a Legendre transformation. The information contained in
the $D(h)$ is the same as that in the $\zeta_j$, and hence
our predicted multifractal spectra obtained
by Legendre transformation are again in good agreement
with the experimental data presented in \cite{boden6}, better
than the predictions of the other models.


Next, let us study the multivariate distribution $p(u_x,u_y,u_z)$
describing the joint probability distribution of the three
components $u_x,u_y,u_z$ of the Lagrangian particle. It is given
by the superstatistical average
\begin{equation}
p(u_x,u_y,u_z)=\frac{1}{(2\pi)^{3/2}}\int_0^\infty
\beta^{3/2}f(\beta)e^{-\frac{1}{2}\beta (u_x^2+u_y^2+u_z^2)}d
\beta .\label{15}
\end{equation}
In particular, for small $\tau$ the distribution of the absolute
value $|\vec{a}|=\tau^{-1}|\vec{u}|$ of acceleration is given by
\begin{eqnarray}
p(|\vec{a}|)&=& 4\pi |\vec{a}|^2 p(a_x,a_y,a_z) \\
&=& \sqrt{\frac{2}{\pi}}|\vec{a}|^2 \tau^3 \int_0^\infty
\beta^{3/2}f(\beta) e^{-\frac{1}{2}\beta \tau^2 |\vec{a}|^2}
d\beta  .\label{9}
\end{eqnarray}
The agreement of this formula with experimentally measured
distributions of the acceleration modulus has been
checked in \cite{boden4}, taking again for $f(\beta )$ the
lognormal distribution. Excellent agreement was found.
Note that in eq.~(\ref{15}) the 3-point density is not the product
of 1-point densities as given by eq.~(\ref{su}), and hence the
model naturally introduces statistical dependence between the
acceleration components.

We may investigate this effect in a quantitative way, by studying
the ratios $R:=p(a_x,a_y)/(p(a_x)p(a_y))$.
From our superstatistical 3-d model we obtain the general
prediction
\begin{equation}
R=\frac{\int_0^\infty \beta
f(\beta)e^{-\frac{1}{2}\beta \tau^2 (a_x^2+a_y^2)}d\beta}{
\int_0^\infty\beta^{1/2}f(\beta )e^{-\frac{1}{2}\beta \tau^2 a_x^2}d\beta
\int_0^\infty\beta^{1/2}f(\beta)e^{-\frac{1}{2}\beta \tau^2 a_y^2}d\beta}
\end{equation}
which is plotted in Fig.~3 for the example of the lognormal
distribution $f(\beta)$.
\begin{figure}[ht]
\includegraphics[scale=0.8]{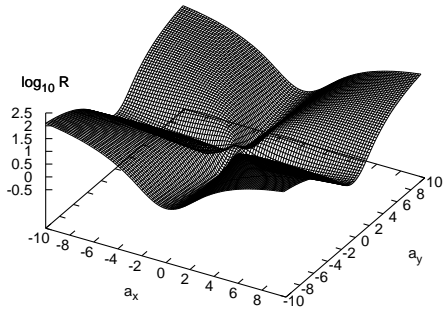}
\caption{The ratio $R=p(a_x,ay)/(p(a_x)p(a_y)$ as predicted by
the superstatistical model  for $s^2=3$. $R\not= 1$ indicates statistical
dependence of acceleration components.}
\end{figure}
One obtains a diagram that closely resembles the experimentally
measured data (Fig.~5 in \cite{boden4}).

All kinds of quantities describing the statistical dependence can
be analytically evaluated for our model. For example, one obtains
the conditional moments $\langle a_x^j|a_y \rangle$ as
\begin{equation}
\langle a_x^j|a_y\rangle =(j-1)!! \tau^{-j} \frac{\int_0^\infty d\beta
\beta^{\frac{1-j}{2}}f(\beta )e^{-\frac{1}{2}\beta
\tau^2 a_y^2}}{\int_0^\infty d\beta \beta^{1/2}f(\beta
)e^{-\frac{1}{2}\beta \tau^2 a_y^2}}
\end{equation}
for $j$ even (they vanish for $j$ odd).
Moreover, for lognormal superstatistics one obtains
\begin{equation}
\frac{\langle a_x^ia_y^j\rangle}{\langle a_x^i\rangle \langle
a_y^j\rangle} =w^{\frac{1}{4}ij}=e^{\frac{1}{4}ijs^2},
\end{equation}
($i,j$ even) which yields a relation between
the flatness parameter $w$ and the statistical
dependencies of the acceleration components that can be checked in future experiments.

Our model also allows for the calculation of temporal correlation
functions. In particular we may be interested in temporal
correlation functions of single components $u_x$ of velocity
differences, i.e.\ $C(t)= \langle u_x(t'+t) u_x(t') \rangle$. By
averaging over the possible random vectors $\vec{n}$ one arrives
at the formula
\begin{equation}
C(t)=\frac{1}{3} \langle u_x^2 \rangle e^{-\gamma t} (2\cos
 B t +1),
\end{equation}
i.e.\ there is rapid (exponential) decay with a zero-crossing at
$t^*=\frac{2}{3}\pi B^{-1}$. Exponential decay and
zero-crossings are indeed observed for the experimental data
\cite{pinton2,pinton3,boden3} as well as in
Lagrangian DNS \cite{yeung3}. The experimentally observed zero
of the correlation function \cite{boden3, pinton2}
can be used to estimate the size of the parameter $B$.
DNS \cite{yeung1,yeung3} indicates that $t^*\approx 2.2\tau_\eta$
independent of $R_\lambda$,
thus our model parameter $B$ is given by $B\approx 0.95 \tau_{\eta}^{-1}$.

Higher-order correlation functions are of interest, too. For
example, the correlation function of the square of velocity
differences $\hat{C} (t)= \langle \vec{u}^2(t'+t) \vec{u}^2(t)
\rangle$ can be approximated as
$\hat{C} (t) \approx \langle \beta^{-1}(t'+t) \beta^{-1}(t')
\rangle$ .
Clearly, by construction of the superstatistical model, this
correlation function decays very slowly, since the process $\beta
(t)$ evolves on a much larger time scale $T_\sigma$ than the
process  $u(t)$. This makes it clear why the correlation function
of the modulus $|\vec{u}|=\sqrt{|\vec{u}|^2}$
decays very slowly, as it is
indeed observed in experiments \cite{boden4,pinton3}
and in DNS \cite{yeung1}.

Finally, let us comment on the typical evolution of dissipation
and enstrophy fluctuations in our model. A large value of local
energy dissipation $\epsilon$ corresponds to a large value of
$\sigma^2$, since in our superstatistical dynamical model
$\epsilon \sim \beta^{-1} \sim \sigma^2/(2\gamma)$. This means
the forcing $\sigma \vec{L}(t)$ is strong for a while. After some
short relaxation time of the order $\gamma^{-1}\sim \tau_\eta$,
where $\tau_\eta$ is the Kolmogorov time, this will create a
large local acceleration variance $\langle a^2 \rangle \sim
\langle u^2 \rangle \tau^{-2} \sim \beta^{-1} \tau^{-2}$. The
term $B\vec{n}\times \vec{u}$ will then create a lot of
rotational energy ($=$enstrophy $\Omega$), as soon as $|\vec{u}|$
has become large. Thus energy dissipation and enstrophy are
strongly correlated in time, and enstrophy lags behind dissipation
evolution by something of the order $\tau_{\eta}$ (the relaxation
time of the system). This is actually experimentally observed in
Fig.~1 of \cite{zeff}. The peaks of $\epsilon$ and $\Omega$ are
shifted by about half a second, which corresponds to the
Kolmogorov time of the system studied by Zeff et al. \cite{zeff}.
Our simple superstatistical model describes these effects in a
correct way. The above time-lag effect also shows up as an
asymmetry of the dissipation-enstrophy cross-correlation function
in DNS \cite{yeung1,yeung3}.

To conclude, we have demonstrated that the most important
statistical phenomena that have been experimentally reported in
LT experiments so far are well reproduced by a
superstatistical model that can be regarded as a generalized
Brownian motion model relevant for 3-d Lagrangian tracer dynamics.
The model arises out of a physically plausible local momentum
balance equation for the Lagrangian particle, and,
compared to other recent Lagrangian models \cite{meneveau},
has the advantage of being analytically treatable.
The model naturally incorporates a
superposition of several stochastic processes, a fast one for
velocity differences of the tracer particle and slow ones for
dissipation and enstrophy in the environment of the tracer
particle. The predicted Lagrangian scaling exponents $\zeta_j$,
the 1-point and 3-point probability densities, correlation
functions, as well as the statistical dependencies between
acceleration components are in excellent agreement with the
experimental data.


\end{document}